\documentclass{IEEEtran}

\usepackage{standalone}

\usepackage{amsmath,graphicx}
\usepackage[utf8]{inputenc}
\usepackage{psfrag,epsfig,graphics}
\usepackage{amsmath,amsthm,amssymb,multirow}
\usepackage{mathbbol}
\usepackage{amssymb}             % AMS Math

\DeclareSymbolFontAlphabet{\amsmathbb}{AMSb}%
\usepackage{mathabx} %      [[  ]] \ldbrack \rdbrack

\usepackage{booktabs}
\usepackage{graphicx}

\usepackage[noadjust]{cite}
\usepackage{multirow}
\usepackage[lined,linesnumbered,ruled]{algorithm2e}

\usepackage{color}  % To include comments in color
\def\cred{\textcolor{red}}

\newcommand{\cb}[1]{{\boldsymbol{#1}}}
\newcommand{\cp}[1]{\ifmmode {\mathcal{#1}}\else ${\mathcal{#1}}$\fi}

\newcommand{\bA}{\boldsymbol{A}}
\newcommand{\bB}{\boldsymbol{B}}
\newcommand{\bC}{\boldsymbol{C}}
\newcommand{\bD}{\boldsymbol{D}}

\newcommand{\bE}{\boldsymbol{E}}
\newcommand{\bF}{\boldsymbol{F}}
\newcommand{\bG}{\boldsymbol{G}}
\newcommand{\bH}{\boldsymbol{H}}
\newcommand{\bI}{\boldsymbol{I}}
\newcommand{\bK}{\boldsymbol{K}}
\newcommand{\bM}{\boldsymbol{M}}

\newcommand{\bP}{\boldsymbol{P}}
\newcommand{\bQ}{\boldsymbol{Q}}
\newcommand{\bR}{\boldsymbol{R}}
\newcommand{\bS}{\boldsymbol{S}}

\newcommand{\bX}{\boldsymbol{X}}
\newcommand{\bY}{\boldsymbol{Y}}

\newcommand{\bm}{\boldsymbol{m}}

\newcommand{\be}{\boldsymbol{e}}

\newcommand{\br}{\boldsymbol{r}}
\newcommand{\by}{\boldsymbol{y}}

\newcommand{\bv}{\boldsymbol{v}}

\usepackage[euler]{textgreek}
\usepackage[colorinlistoftodos]{todonotes}

\newcommand{\bpsi}{\boldsymbol{\psi}}

\newcommand{\btheta}{\boldsymbol{\theta}}

\newcommand{\bDelta}{\boldsymbol{\Delta}}

\newcommand{\bPsi}{\boldsymbol{\Psi}}
\newcommand{\bSigma}{\boldsymbol{\Sigma}}

\newcommand{\tr}{\operatorname{tr}}
\newcommand{\Ex}{\operatorname{E}}
\newcommand{\vect}{\operatorname{vec}}
\newcommand{\diag}{\operatorname{diag}}

\usepackage{color}  % To include comments in color
\def\cred{}

\definecolor{darkgreen}{rgb}{0., 0.4, 0.}
\definecolor{amber}{rgb}{1.0, 0.49, 0.0}
\definecolor{orange}{rgb}{1.0, 0.4, 0.0}

\usepackage[bibbreaks=tight,
            paragraphs=tight,
            floats=normal,
            mathspacing=normal,
            wordspacing=normal,
            tracking=normal,
            bibnotes=tight,
            charwidths=normal,
            mathdisplays=normal,
            leading=normal,
            indent=normal,
            lists=normal,
            bibliography=normal,
            title=normal,
            sections=normal,
            margins=normal,
            ]{savetrees}

\title{Kalman Filtering and Expectation Maximization for\\ Multitemporal Spectral Unmixing}

\author{Ricardo A. Borsoi,~~Tales Imbiriba,~~Pau Closas,~~José C.~M. Bermudez,~~C\'edric Richard

\thanks{
R.A.Borsoi is with Dept. of Electrical Engineering, Universidade Federal de Santa Catarina (DEE-UFSC), Florian\'opolis, Brazil, and with the Lagrange Laboratory, Universit\'e C\^ote d'Azur (UCA), Nice, France.

T.Imbiriba and P.Closas are with the Dept. of Electrical \& Computer Engineering, Northeastern University, Boston, MA, USA.

J.C.M.Bermudez is with DEE-UFSC and with the Dept. of Electronic Engineering and Computing, Catholic University of Pelotas, Pelotas, Brazil.

C.Richard is with the Lagrange Laboratory, UCA, Nice, France.

This work has been supported by the National Council for Scientific and Technological Development (CNPq), and the National Science Foundation under Awards CNS-1815349 and ECCS-1845833.}}

\allowdisplaybreaks
\begin{document}
%\ninept
%

\maketitle
\begin{abstract}
% The recent evolution of hyperspectral imaging technology and the proliferation of new emerging applications presses for the processing of multiple temporal hyperspectral images. In this work, we propose a novel spectral unmixing (SU) strategy which accounts for temporal spectral variability. Specifically, we re-write the multi-temporal mixing process using a state-space formulation for the endmember variability coefficients allowing us to exploit the Bayesian filtering machinery and the expectation maximization (EM) algorithm to estimate other model parameters. The resulting iterative unmixing algorithm is then obtained by combining a Kalman smoother with the EM under the consideration of constant abundances over a short-time window. Simulations show prominent results indicating that the proposed strategy is promising, and deserves to be used and further developed in the context of new multitemporal SU applications. 

The recent evolution of hyperspectral imaging technology and the proliferation of new emerging applications presses for the processing of multiple temporal hyperspectral images. In this work, we propose a novel spectral unmixing (SU) strategy using physically motivated parametric endmember representations to account for temporal spectral variability. By representing the multitemporal mixing process using a state-space formulation, we are able to exploit the Bayesian filtering machinery to estimate the endmember variability coefficients. Moreover, by assuming that the temporal variability of the abundances is small over short intervals, an efficient implementation of the expectation maximization (EM) algorithm is employed to estimate the abundances and the other model parameters. Simulation results indicate that the proposed strategy outperforms state-of-the-art multitemporal SU algorithms. 
\end{abstract}
\begin{keywords}
% Hyperspectral data, endmember variability, multitemporal unmixing, Kalman filter, expectation maximization.
Hyperspectral, endmember variability, multitemporal unmixing, Kalman filter, expectation maximization.
\end{keywords}
\section{Introduction}
\label{sec:intro}

% Hyperspectral imaging is currently at the core of a large and increasing number of applications due to its ability to provide meaningful information about the distribution of the constituent materials across a given scene~\cite{Bioucas-Dias-2013-ID307}. However, due to the low resolution of typical hyperspectral images (HI), the image region captured by each HI pixel usually contains distinct materials.
%
Spectral Unmixing (SU) aims to decompose a hyperspectral image (HI) into its pure spectral components, termed \emph{endmembers} (EMEs), and the proportional \emph{abundances} to which they contribute to the reflectance in each pixel~\cite{Keshava:2002p5667}. Although the interaction between light and the EMEs can be complex and nonlinear~\cite{Dobigeon-2014-ID322, Imbiriba2016_tip, borsoi2019blind}, the linear mixing model (LMM) is still widely used due to its simplicity and good performance~\cite{Keshava:2002p5667}.

Spectral variability (SV) consists of changes in EME spectra occurring both within a single image, or between images acquired at different time instants. They can be caused by differences in atmospheric, illumination or seasonal conditions~\cite{Zare-2014-ID324-variabilityReview,somers2011variabilityReview}. Early approaches have considered large libraries of spectral signatures to represent variable EME spectra~\cite{Zare-2014-ID324-variabilityReview,roberts1998originalMESMA,Borsoi_multiscale_lgrs_2018,borsoi2019deepGenSpectralLibrary}. More recently, different extensions of the LMM have been proposed to account for the spectral variability within a given HI, by considering, e.g., additive~\cite{Thouvenin_IEEE_TSP_2016_PLMM} and multiplicative~\cite{drumetz2016blindUnmixingELMM, imbiriba2018GLMM, borsoi2019improved} scaling factors, or by parametrizing spectral variability using deep generative models~\cite{borsoi2019deep}.

Multitemporal SU have recently become a subject of great interest due to the possibility of leveraging time information in HI sequences, allowing for monitoring the dynamical evolution of the materials and their distributions~\cite{somers2013invasiveHawaiiMultiTemporalBandWeighting,lippitt2018multidateMESMAshrublands,somers2013uncorrelatedBandSelectionInstabilityIndex}. However, the influence of spectral variability in multitemporal scenarios can be significantly stronger than in the case of a single HI. This introduces a challenge to multitemporal SU since EME variability must be carefully modelled to achieve a good performance.
Previous work have considered different strategies to incorporate dynamical information about the EMEs, often based on parametric models originally devised to account for variations within a single HI. These include constraining the EMEs in adjacent time instants to be scaled versions of each other~\cite{henrot2016dynamical}, or to be represented as a mean EME matrix with small, additive perturbations~\cite{thouvenin2016online,sigurdsson2017sparseDistU,thouvenin2018hierarchicalBU}.
However, these works disregard important information as they do not account for the low-dimensional structure that often underlies the changes observed in EME spectra when representing its evolution.

In this paper we propose a new algorithm for multitemporal SU which is based on  a dynamical model for the EME time variability. \cred{Specifically, we couple the representation power of recently proposed parametric EME models (which were originally devised to operate within a single HI, such as~\cite{imbiriba2018GLMM}) with a Bayesian filtering methodology to reliably estimate the endmembers in HI sequences.} Instead of operating directly on the EME spectral space, we make use of a parametric EME model to represent EME dynamics indirectly through vectors of parameters that capture the time variations of each material. Bayesian filtering and smoothing are combined with the expectation maximization algorithm to estimate the required parameters given a window of observations in time. The initialization of the resulting Kalman filter is also estimated in the process, which improves convergence for short image sequences. Under some approximations about the temporal variation of the abundances, the proposed algorithm is able to blindly estimate the EMEs, the average abundances, and the remaining model parameters from the observed HI data. Finally, a unique abundance matrix is estimated for each time instant using the resulting EME model. Simulation results show that, for small abundance variations over time (which can be usually satisfied in small time windows), the proposed method is able to outperform state-of-the-art algorithms in both EME and abundance estimation accuracy.

    % In this paper, we consider a different dynamical strategy to deal with SV with MTHIs within two main assumptions: 1) That for a time window $T$ the changes in the abundances are negligible, and 2) that the we have access to all images within this window.
    % Under such conditions we propose to expand the GLMM by considering a state-space model allowing us to exploit the Bayesian filtering and smoothing machinery while the consideration of negligible abundance variation allows for considering it as model parameters which can be estimated together with other model parameters. Specifically, we propose to consider the expectation maximization strategy to estimate such parameters. 

% This paper is organized as follows. Section~\ref{sec:MTSU} presents the multitemporal SU problem and a more in-depth discussion of related works. Section~\ref{sec:dynModel} presents the proposed dynamical model while Section~\ref{sec:proposedMethod} presents the proposed estimation methodology. Section~\ref{sec:Simulations} presents the simulations results and Section~\ref{sec:conclusions} our final remarks.

% \cblue{[Propose to make Sec 2 a subsection of the introduction and to shrink the introduction as much as possible!]}

% \vspace{-0.8cm}
% \clearpage
% ------------------------------------------------
\section{Multitemporal Spectral Unmixing}\label{sec:MTSU}

The multitemporal Linear Mixing Model (LMM)~\cite{Keshava:2002p5667} represents an HI with $L$ bands and $N$ pixels at time $t$ as:
%
% \vspace{-0.15cm}
\begin{align} \label{eq:LMM}
 &\bY_{\!t} = \bM_t \bA_t + \bE_t, %\\&
\,\,\, \text{subject to }\,\cb{1}^\top\bA_t = \cb{1}^\top,\,\,\, \bA_t \geq \cb{0} \,,%\nonumber
\end{align}
% \vspace{-0.5cm}

% \noindent where $\bY_{\!t}\in\amsmathbb{R}^{L\times N}$ is the observed HI at time $t\in\{1,\ldots,T\}$, the columns of matrix $\bM_t \in \amsmathbb{R}^{L\times P}$ are the $P$ endmember spectral signatures at instant $t$, $\bA_t \in \amsmathbb{R}^{P\times N}$ is a matrix containing the abundances for each pixel, and $\bE_t$ represents additive noise.

\noindent \cred{where $\bY_{\!t}\in\amsmathbb{R}^{L\times N}$ is the observed HI, the columns of $\bM_t \in \amsmathbb{R}^{L\times P}$ are the $P$ endmember spectral signatures, $\bA_t \in \amsmathbb{R}^{P\times N}$ contains the abundances for each pixel, and $\bE_t$ represents additive noise, all indexed at time $t\in\{1,\ldots,T\}$.}

An important challenge related to the use of representation~\eqref{eq:LMM} regards the consideration of spectral variability, which causes the signatures of the endmembers in $\bM_t$ to change due to, e.g., seasonal, illumination or acquisition variations~\cite{Zare-2014-ID324-variabilityReview}. Spectral variability occurs both in space (within the same HI) and in time. Spatial domain spectral variability has been addressed in several works (e.g., see~\cite{Zare-2014-ID324-variabilityReview,somers2011variabilityReview,Thouvenin_IEEE_TSP_2016_PLMM,drumetz2016blindUnmixingELMM,imbiriba2018GLMM,borsoi2019improved,borsoi2019deep} and references therein). For simplicity, this work assumes only variations of EMEs in time.
% EME variation withing the same HI can be later incorporated to the proposed model using, for instance, the compatible approach proposed in [REFERENCE].}
EME variation within the same HI can be later incorporated to the proposed model, for instance, by adapting models such as the one in~\cite{imbiriba2018GLMM} to represent the space-time dynamical behavior of the EMEs.

%An important challenge related to model~\eqref{eq:LMM} consists of spectral variability, which causes the signatures of the endmembers in $\bM_t$ to change over time due to, e.g., seasonal, illumination or acquisition variations~\cite{Zare-2014-ID324-variabilityReview}. 

% \todo[inline]{Note that spectral variability occurs both in time and in space (within the same time sample). The spatial variability is not been considered in this work, but we cannot pretend it does not exist.}

%
% A straightforward way to perform SU under time variability is to do it for each image separately. Doing this, however, disregards the dynamics of the temporal variations, what leads to suboptimal performance.

A straightforward way to perform SU under time variability is to do it for each image separately. \cred{However, such an approach disregards the temporal information and the time dynamics of the spectral variability, which can be exploited to enhance both the abundance and EME estimation performance.}
Different SU algorithms accounting for endmember time variability have been recently proposed, most of them inspired by models designed to account for spectral variability within a single image.
%
%Recently, different SU algorithms were proposed by incorporating dynamical information about the endmembers, mostly inspired by models that account for spectral variability within a single image. 
For instance, in~\cite{henrot2016dynamical} the authors constrain the endmember matrices at each time instant to be scaled versions of a reference endmember matrix. In~\cite{thouvenin2016online}, the authors model the endmembers at each time instant by a mean EME matrix plus small perturbations, which are assumed to be temporally smooth. All variables are then estimated using a stochastic approach. This latter model was later extended for distributed unmixing with additional sparsity constraints in~\cite{sigurdsson2017sparseDistU}, and to include sparse additive residual terms to represent abrupt spectral variations in the HI using a hierarchical Bayesian framework in~\cite{thouvenin2018hierarchicalBU}. 
%
%
% However, these works do not provide a satisfactory means of modeling the dynamical evolution of the endmembers since \cblue{they do not take into account the low-dimensional structure that often underlies the changes observed in EME spectra.}
However, these works do not provide a satisfactory means of modeling the dynamical evolution of the endmembers since they operate directly in the input spectral space, ignoring the fact that spectral variability can often be represented more accurately using physically meaningful parametrizations of EME spectra.

% \todo[inline]{What do you mean by ``low dimensional structure'' in this case? If it is the number of parameters used to parametrize the variability, we use $LP$ parameters in $\bpsi_t$, which is the same number of parameters used in~\cite{thouvenin2016online}, or not? If it is the smoothness of the time variations, the model \eqref{eq:param_state_evolution} is a random walk model, and does not impose a smooth variation of $\bpsi_t$. If matrix $\bQ$ is full, we are modeling spectral correlation, not time correlation, in $\bpsi_t$. A time smoothness would result from using a system matrix $\bF$ different from identity, like $\bpsi_{t} = \bF\bpsi_{t-1} + \cb{q}_{t}$.}

Different models have been recently proposed to model EME spatial variability as a parametric function of reference spectral signatures as:
% \vspace{-0.15cm}
\begin{align} \label{eq:parametric_EM_model}
    % \bM_{t+1} = g(\bM_0,\bpsi_{t})
    \bM = f(\bM_0,\bpsi) \,,
\end{align}
% \vspace{-0.5cm} \noindent 
where $f$ is a parametric function, $\bM_0\in\amsmathbb{R}^{L\times P}$ contains reference/average spectral signatures and $\bpsi$ is a vector of parameters of the variability model.
Such models include additive perturbations~\cite{Thouvenin_IEEE_TSP_2016_PLMM}, spectrally uniform~\cite{drumetz2016blindUnmixingELMM} or spectrally varying~\cite{imbiriba2018GLMM, borsoi2019improved}, and parametrizations using deep neural networks learned from the observed HI~\cite{borsoi2019deep}.
%\cred{The parametric models used in \cite{drumetz2016blindUnmixingELMM,imbiriba2018GLMM, borsoi2019improved} are specially interesting for building a dynamical model to consider EME time variability.}
Such parametric models are specially interesting for building a dynamical model to consider EME time variability.
%
% Such models are able to capture the parsimony of the endmembers' factors of variation, and have been very successful in representing spectral variability within HIs in both unmixing and image fusion problems. 
%
% In the following, we explore such models in order to obtain a parsimonious representation of endmember dynamics in a multitemporal setting.

% \begin{color}{blue}
% The LMM assumes that the endmember spectra are fixed for all pixels $\br_n$, $n=1,\ldots,N$, in the HI. This assumption jeopardizes the accuracy of estimated abundances in many circumstances due to the spectral variability existing in a typical scene.

% To mitigate this issue, the GLMM~\cite{imbiriba2018GLMM} models arbitrary EM variability by considering band-dependent multiplicative factors, which links the amount of spectral variability to the magnitude of the EM reflectance in each band.

% The GLMM introduces a scaling matrix $\cb{\Psi}_n\in{\amsmathbb{R}^{L\times R}}$ with entries $[\cb{\Psi}_n]_{\ell,k} = \psi_{n_{\ell,k}}\geq 0$, which acts individually on each wavelength. The model represents the $n$-th HI pixel as
% \begin{equation} \label{eq:model_glmm}
%     \br_n = (\bM\odot\cb{\Psi}_n)\bA_t + \be_n
% \end{equation}
% where $\odot$ stands for the Hadamard product.
% \end{color}

% \vspace{-0.2cm}
% ------------------------------------------------
\section{Dynamical parametric endmember model}\label{sec:dynModel}
% \vspace{-0.1cm}

In this paper, we consider a multitemporal extension of the parametric EME model~\eqref{eq:parametric_EM_model}. We assume a fixed reference EME matrix $\bM_0$, and model the time variations in $\bM_t$ through a time varying $\bpsi_t$, $t=1,\ldots,T$. 
By assuming that temporally adjacent images are acquired at reasonably short time intervals, we model the difference $\bpsi_{t} - \bpsi_{t-1}$ as a small zero-mean vector. Thus, we assume the following model for~$\bpsi_t$:
% \vspace{-0.1cm}
\begin{align} \label{eq:param_state_evolution}
    \bpsi_{t} &= \bpsi_{t-1} + \cb{q}_{t}
\end{align}
where $\bpsi_{t}$ is a vector containing the parameters of the endmember model at time $t$, and $\cb{q}_{t}\sim\mathcal{N}(\cb{0},\bQ)$ contains the innovations which describe its dynamical evolution.
Note that $\cb{q}_{t}$ is only constrained to be zero mean on statistical and not temporal average, which means that each realization of the sequence $\{\widehat{\cb{q}}_{t}\}$, which is learned from the observed HIs, can exhibit behavior such as trends and complex dynamic evolution. Moreover, the Gaussian assumption is only made in the model parameters $\bpsi_{t}$ and not on the EME signatures themselves, which allows for the use of complex EME distributions through the pushforward measure obtained using the function~$f$, as done in~\cite{borsoi2019deep}.
This generalizes parametric EME model~\eqref{eq:parametric_EM_model} to the multitemporal setting as $\bM_{t} = f(\bM_0,\bpsi_{t})$, where the parametric function $f$ now relates the EME matrices and the vectors of parameters at each time instant.
Considering this model, the multitemporal LMM can be represented as
% \vspace{-0.1cm}
\begin{align} \label{eq:obs_model_general}
    \bY_{\!t} = f(\bM_0,\bpsi_{t})\bA_t + \bE_t \,.
\end{align}
% \vspace{-0.555cm}

Next, one must choose a function $f$ for \eqref{eq:obs_model_general} that establishes a good compromise between mathematical tractability and performance.  The GLMM~\cite{imbiriba2018GLMM,Borsoi_2018_Fusion} is able to represent arbitrary spectral variability by considering spectrally varying multiplicative scaling factors, introducing a connection between the amount of spectral variability and the amplitude of endmember reflectance spectra at each band.
The GLMM introduces a matrix $\cb{\Psi}\in{\amsmathbb{R}^{L\times P}}$ of scaling factors with nonnegative entries $[\cb{\Psi}]_{\ell,k} \geq 0$ acting individually at each wavelength. This leads to the following representation for the $t$-th observed HI:
\begin{equation} \label{eq:model_glmm}
    \bY_{\!t} = (\bM_0\odot\cb{\Psi}_t)\bA_t + \bE_t \,,
\end{equation}
where $\odot$ is the Hadamard (elementwise) product.
Using the vectorization property,~\eqref{eq:model_glmm} can be expressed as
\begin{align}
    \by_t & = \vect(\bY_{\!t})= \big(\bA_t^\top\otimes\bI_L\big)\diag(\bm_{0}) \bpsi_{t} + \be_{t} \,,
\end{align}
with $\bm_{0} = \vect(\bM_0)$, $\bpsi_{t} = \vect(\bPsi_{t})$ and $\be_t = \vect(\bE_t)$.

We write the abundance matrix $\bA_t$ as
%Suppose that for a time window $t\in \{1,\ldots, T\}$, the abundances can be written as
$
\bA_t = \bA + \bDelta\!\bA_t
$
% \begin{align}
%     \bA_t = \bA + \bDelta\!\bA_t
% \end{align}
where $\bDelta\bA_t$ represents small random fluctuations over the average abundance matrix $\bA$. Considering $\bDelta\!\bA_t$ to be small for a time window $t\in \{t_0,\ldots, t_0+T\}$, $\forall t_0$, 
%where $\bDelta\bA_t$ represents small random fluctuations over the average abundance matrix $\bA$. If we consider $\bDelta\!\bA_t$ to be small, we can incorporate 
these variations can be incorporated into the observation noise, leading to the following model:
\begin{align} \label{eq:observation_model}
    \by_t & = \bH(\bA)\diag(\bm_{0}) \bpsi_{t} + \br_{t} \,,
\end{align}
where $\bH(\bA)=\bA^\top\otimes\bI_L$ and $\br_{t}=\be_{t}+\big(\bDelta\!\bA_t^\top\otimes\bI_L\big)\diag(\bm_{0}) \bpsi_{t}$. 
Note that the observation noise $\br_{t}$ in~\eqref{eq:observation_model} is correlated with the state $\bpsi_{t}$. \cred{In the following, we will use a signal-independent noise approximation, which provides competitive performance at a modest computational cost. Further discussion on the impact of such an approximation can be found in the supplementary document, also available in~\cite{borsoi2020kalmanEM_arxiv}.}

\section{Proposed method}\label{sec:proposedMethod}
% \vspace{-0.1cm}

In this section, we present the proposed dynamical methodology which connects Kalman smoother with expectation maximization approach. For this, we assume that for a given time window of duration $T$ the abundance variation is small but the EMEs can vary due to different seasonal or acquisitions conditions. Then, we employ a time varying state space formulation to model the spectral variability, which naturally leads to a Kalman filter based formulation.
We couple a Kalman smoother, used to obtain accurate estimations for the state variables, with the Expectation Maximization estimation of model parameters such as the abundance matrix and the noise power. 
Assuming the abundances fixed over a time window $t\in \{t_0,\ldots, t_0+T\}$, we use~\eqref{eq:param_state_evolution} and~\eqref{eq:observation_model} to form the linear state-space model
% \vspace{-0.15cm}
\begin{align}
\begin{split} \label{eq:state_space_model}
    \bpsi_{t} &= \bpsi_{t-1} + \cb{q}_{t},
    % \by_{t} &= \bH(\btheta)vec(\bM_0\odot\bPsi_{t}) + \br_{t} \nonumber\\
    % &= \bH(\btheta)vec(\bM_0)\odot vec(\bPsi_{t}) + \br_{t} \nonumber\\
    \quad \by_{t} = \bH(\bA)\diag(\bm_{0}) \bpsi_{t} + \br_{t} \,.
\end{split}
\end{align}
% \vspace{-0.5cm}

Neglecting the dependence of $\br_t$ on $\bpsi_{t}$, and assuming $\cb{q}_t$ and $\br_t$ to be Gaussian, this system can be solved using the classical Kalman filter and smoothing equations.
%
%\noindent Since the above state-space model is linear with respect to both $\bA$ and $\bpsi_t$, and we assume $\cb{q}_t$ and $\br_t$ to be Gaussian, the Bayesian filter and smoother boils down to the well known Kalman filter and smoothing equations.
Next, we present the Kalman filter and smoother equations followed by the EM strategy to estimate the abundances and noise power. 
% It is important to 
% with $\bpsi_{t} = vec(\bPsi_{t})$, $\bm_{0} = vec(\bM_0)$.

% \cblue{Descriptive text here: In this approach we consider abundances as model parameters while considering $\bpsi$'s as states changing over time. Assuming that during a time window of length $T$  we have constant abundances we propose to consider a Bayesian smoother/EM to model the dynamic abundance and estimate the abundances...}

% \vspace{-0.3cm}
\subsection{Kalman Filter endmember model}
\label{sec:kalmanFilt}
Bayesian filtering computes marginal posterior distributions of the states by assuming Markovity over the state sequence. 
When the dynamical and measurement models are linear and Gaussian, the solution is given in the form of the Kalman filter, which can be expressed in a set of equations for the~Prediction:
\begin{align} \label{eq:kalman_filt_pred} 
\begin{split}
    \bpsi_{t|t-1} = \bpsi_{t-1|t-1}\,, \hspace{6ex}
    \bP_{t|t-1} = \bP_{t-1|t-1} + \bQ \,,
\end{split}
\end{align}
% \vspace{-0.5cm}

% \noindent Update step:
% \begin{align}
%     \bv_t &= \by_t - \bB\bpsi_{t|t-1} \nonumber\\
%     \bS_{t} &=  \bB \bP_{t|t-1} \bB^\top + \bR \nonumber\\
%     \bK_t &= \bP_{t|t-1} \bB^\top \bS_{t}^{-1} \label{kalman_filt_update}\\
%     \bpsi_{t|t} &= \bpsi_{t|t-1} + \bK_t \bv_t  \nonumber\\
%     \bP_{t|t} &= \bP_{t|t-1} - \bK_t \bS_{t} \bK_t^\top  \nonumber
% \end{align}

\noindent and for the Update step:\\
\noindent\begin{minipage}{\linewidth}
\vspace{-5pt}
\noindent\begin{minipage}{.4\linewidth}
\begin{align}
    \bv_t &= \by_t - \bB\bpsi_{t|t-1} \,,\nonumber\\
    \bS_{t} &=  \bB \bP_{t|t-1} \bB^\top + \bR \,,\nonumber\\
    \bK_t &= \bP_{t|t-1} \bB^\top \bS_{t}^{-1} \,,\nonumber
\end{align}
\end{minipage}
\noindent\begin{minipage}{.6\linewidth}
\begin{align}\label{kalman_filt_update}
\begin{split}
    \bpsi_{t|t} &= \bpsi_{t|t-1} + \bK_t \bv_t \,,\\
    \bP_{t|t} &= \bP_{t|t-1} - \bK_t \bS_{t} \bK_t^\top \,,% \nonumber
\end{split}
\end{align}
\end{minipage}
\vspace{5pt}
\end{minipage}
where $\bP_{t_1|t_2}$ is the covariance matrix of $\bpsi_{t_1}$ conditioned on $\by_t$ for $t_1 \le t_2$, $\bB = \bH(\bA)\diag(\bm_{0}) =(\bA^\top \otimes \bI_L) \diag(\bm_{0})$ and $\bR$ is the covariance matrix of $\br_t$ in \eqref{eq:state_space_model}.
Solving equations~\eqref{kalman_filt_update} requires to construct and invert matrix $\bS_t$ of size $NL\times NL$, which is impractical.
To circumvent this issue, we assume that the noise covariance matrix satisfies $\bR=\sigma_r^2\bI_{NL}$. Thus, using the Woodbury identity for the inverse of sum of matrices, the right part of the third term in~\eqref{kalman_filt_update} is written as
\begin{align}
    \bB^\top \bS_{t}^{-1} 
    %
    % & = \bB^\top \big(\bB \bP_{t|t-1} \bB^\top + \bR\big)^{-1}
    % \\
    % & = \bB^\top \bR^{-1} - \bB^\top \bR^{-1} \bB \big(\bP_{t|t-1}^{-1} + \bB^\top \bR^{-1} \bB \big)^{-1} \bB^\top \bR^{-1} \nonumber
    % \\
    & = \sigma_r^{-2}\bB^\top - \sigma_r^{-4}\bB^\top \bB \big(\bP_{t|t-1}^{-1} + \sigma_r^{-2}\bB^\top \bB \big)^{-1} \bB^\top \nonumber
\end{align}
which now involves only the inverse of a $PL\times PL$ matrix.

% One issue with the Kalman filter equations presented above is the necessity to construct and invert matrices of size $NL\times NL$, which is impracticable. Possible ways to circumvent this issue include exploiting the simplified structure often considered for $\bR$. 

% Thus, using the Woodbury identity for the inverse of sum of matrices, we have:
% \begin{align}
%     \bB^\top \bS_{t}^{-1} & = \bB^\top \big(\bB \bP_{t|t-1} \bB^\top + \bR\big)^{-1}
%     \\
%     & = \bB^\top \bR^{-1} - \bB^\top \bR^{-1} \bB \big(\bP_{t|t-1}^{-1} + \bB^\top \bR^{-1} \bB \big)^{-1} \bB^\top \bR^{-1} \nonumber
% \end{align}

% Assuming $\bR = \sigma_r^2 \bI_{NL} \Rightarrow \bR^{-1} = \sigma_r^{-2} \bI_{NL}$ the terms in the Woodbury identity becomes:
% \begin{itemize}
%     \item $\bB^\top \bR^{-1} = \sigma_r^{-2}\bB^\top$
%     \item $\bB^\top \bR^{-1} \bB = \sigma_r^{-2}\bB^\top\bB$ which is a $PL\times PL$ matrix and thus feasible. 
% \end{itemize}
% % \cblue{We could further explore the sparse structure of $\bB$ for saving memory.}

% \vspace{-0.3cm}
\subsection{Kalman Smoother}
\label{sec:kalmanSmooth}

% The purpose of Bayesian smoothing is to access the marginal posterior distribution of state $\bpsi^{(t)}$ after receiving all
The objective of Bayesian smoothers is to provide a marginal posterior distribution of state $\bpsi_t$ assuming knowledge of the measurements $\by_{t}$ 
% \cred{in an obseration window of duration $T$, that is, $p(\bpsi_t|\by_T)$, where $\by_T = [y_{t_0}, \dots, y_{t_0+T}]^\top$, $\forall t_0$.}
in an observation window of duration $T$, that is, $p(\bpsi_t|\by_{t_0},\ldots,\by_{t_0+T})$.
% \begin{align}, 
%     p\left(\bpsi_t|\by_{1:T}\right).
% \end{align}
% \cred{[explain the smoother a bit better here]} For the model in~\eqref{eq:state_space_model}, the smoother equations are given by:
For the model in~\eqref{eq:state_space_model}, the smoother solution can be implemented very efficiently by iteratively updating the conditional distributions obtained by the Kalman filter backwards in time, for $t_0+T,\ldots,t_0$. In this case, the smoother equations are given by:
% \vspace{-0.4cm}
% \begin{align} \label{eq:kalman_smoother}
%     \bpsi_{t+1|t} &= \bpsi_{t|t} \nonumber\\
%     \bP_{t+1|t} &= \bP_{t|t} + \bQ  \nonumber\\
%     \bG_t &= \bP_{t|t} \bP_{t+1|t}^{-1}\\
%     \bpsi_{t}^{s} &= \bpsi_{t|t} + \bG_t [\bpsi_{t+1}^{s} - \bpsi_{t+1|t}]  \nonumber\\
%     \bP^s_{t} &= \bP_{t|t} + \bG_t [\bP^s_{t+1} - \bP_{t+1|t}]\bG_t^\top. \nonumber
% \end{align}
%
% \begin{align} \nonumber
%     \bpsi_{t+1|t} = \bpsi_{t|t} \,, \hspace{3ex}
%     \bP_{t+1|t} = \bP_{t|t} + \bQ \,, \hspace{3ex}
%     \bG_t = \bP_{t|t} \bP_{t+1|t}^{-1}
% \end{align}
% \vspace{-15pt}
% \begin{align} \label{eq:kalman_smoother}
% \begin{split}
%     \bpsi_{t}^{s} &= \bpsi_{t|t} + \bG_t [\bpsi_{t+1}^{s} - \bpsi_{t+1|t}] \,,  \\
%     \bP^s_{t} &= \bP_{t|t} + \bG_t [\bP^s_{t+1} - \bP_{t+1|t}]\bG_t^\top. 
% \end{split}
% \end{align}
\begin{align} \label{eq:kalman_smoother}
\begin{split}
    \bpsi_{t+1|t} = \bpsi_{t|t} \,,  &  \hspace{3ex}
    \bP_{t+1|t} = \bP_{t|t} + \bQ \,, \hspace{3ex}
    \bG_t = \bP_{t|t} \bP_{t+1|t}^{-1} %\,,
\end{split}
\nonumber \\
\begin{split}
    \bpsi_{t}^{s} &= \bpsi_{t|t} + \bG_t [\bpsi_{t+1}^{s} - \bpsi_{t+1|t}] \,,  \\
    \bP^s_{t} &= \bP_{t|t} + \bG_t [\bP^s_{t+1} - \bP_{t+1|t}]\bG_t^\top.
\end{split}
\end{align}

% \vspace{-0.4cm}
% -------------------------------------------------------------------------------
\subsection{The expectation-maximization (EM) algorithm}
\label{sec:expectM}

Estimation of the sequence $\{\bpsi_t\}$ of EME model parameters using \eqref{eq:kalman_filt_pred}-\eqref{eq:kalman_smoother}, requires that $\bA$, $\bQ$ and $\bR$, as well as the initializations $\bP_{0|0}$ and $\bpsi_{0|0}$, be known in advance.
Let us denote these parameters by $\btheta=\{\bA,~\bP_{0|0},~\bQ,~\bR,~\bpsi_{0|0}\}$. Instead of fixing $\btheta$ with values known a priori, we can view it as unobserved latent variables of model~\eqref{eq:state_space_model}, which can be estimated by maximizing the conditional marginal likelihood
%~\cred{$p(\by_T|\btheta)$} 
$p(\by_{t_0},\ldots,\by_{t_0+T}|\btheta)$
using the EM algorithm.
Starting with an initial guess $\btheta^{(0)}$, the EM algorithm finds a local maximum of~$p(\by_{t_0},\ldots,\by_{t_0+T}|\btheta)$ by repeating the following steps:
% \vspace{-5pt}
\begin{align} \label{eq:EM_alg_iters}
\begin{split}
    \quad & a) \,\,\, \text{E-step: compute } \mathcal{Q}(\btheta|\btheta^{(k)})
    \\
    \quad & b) \,\,\, \text{M-step: compute }  \btheta^{(k+1)}=\arg\max\nolimits_{\btheta}\mathcal{Q}(\btheta,\btheta^{(k)})
\end{split}
\end{align}

\vspace{-3pt}
% \begin{itemize}
%     \item E-step: compute $\mathcal{Q}(\btheta|\btheta^{(k)})$
%     \item M-step: compute $\btheta^{(k+1)}=\arg\max_{\btheta}\mathcal{Q}(\btheta,\btheta^{(k)})$
% \end{itemize}
\noindent for $k=1,\ldots,K_{max}$, with $K_{max}$ being the number of iterations and
% \cred{$\mathcal{Q}(\btheta|\btheta^{(k)})=\Ex_{p(\bpsi_{T},\by_{T}|\btheta^{(k)})}\{\log p(\bpsi_{T},\by_{T}|\btheta)\}$, with $\bpsi_{T} = [\bpsi_{t_0}, \dots, \bpsi_{t_0+T}]^\top$,}
$\mathcal{Q}(\btheta|\btheta^{(k)})=\Ex_{\varsigma}\{\log p(\bpsi_{t_0}, \ldots, \bpsi_{t_0+T},\by_{t_0},\ldots, \by_{t_0+T}|\btheta)\}$, with $\varsigma=p(\bpsi_{t_0}, \ldots,\bpsi_{t_0+T},\by_{t_0},\ldots,\by_{t_0+T}|\btheta^{(k)})$, 
being the expectation of the logarithm of the data likelihood, taken with respect to the full joint posterior given the parameters~$\btheta^{(k)}$.
Although the EM algorithm is very general and not always easy to solve, for a linear model such as~\eqref{eq:state_space_model}  we can find closed form solutions, leading to a more efficient implementation in high-dimensional settings. Furthermore, for the linear Gaussian model, $\mathcal{Q}(\btheta,\btheta^{(k)})$ can be computed based on the Kalman smoother results obtained using $\btheta^{(k)}$ as the system parameters. This leads to an elegant solution that consists of the successive application of the smoother and estimation of the parameters.
For the model~\eqref{eq:state_space_model}, $\mathcal{Q}(\btheta,\btheta^{(t)})$ is given by~\cite{sarkka2013bayesian}
\begin{align}% \label{eq:Q}
    & \mathcal{Q} (\btheta,\btheta^{(t)}) \!=\! -\frac{1}{2}\Big(\!
    \tr\!\Big\{\!\bP_{0|0}^{-1}\big[ \bP_0^s \!+ \!(\bpsi_0^s\!-\!\bpsi_{0|0})(\bpsi_0^s\!-\!\bpsi_{0|0})^\top \big]\Big\}
    \nonumber \\
    & + \! \tr\!\Big\{\!\bR^{-1}\big[ \bSigma_5 - \bSigma_3 \bH(\!\bA)^\top 
    \!\! - \! \bH(\!\bA) \bSigma_3^\top  + \bH(\!\bA) \bSigma_1 \bH(\!\bA)^\top \big]\Big\}
    \nonumber \\
    & + \! \tr\!\Big\{\!\bQ^{-1}\big[ \bSigma_1 \!-\! \bSigma_4 \!-\! \bSigma_4^\top \!+\! \bSigma_2 \big]\Big\} \!+\!\log\!\big\{|\bQ\bR|^T|\bP_{0|0}|\big\} \!\Big) \!+\!C
    \nonumber
\end{align}
where $C$ is a constant term and 
% \vspace{-0.1cm}
% \begin{align} \nonumber 
%     \bSigma_1  = \sum_{t=1}^T \bP_{t}^s + \bpsi_{t}^s {\bpsi_{t}^s}^\top,
%     \hspace{4.25ex}
%     \bSigma_4  =  \sum_{t=1}^T \bP_{t}^s {\bG^s}^\top_{t-1} + \bpsi_{t}^s {\bpsi^s}^\top_{t-1}
% \end{align}
% \vspace{-0.5cm}
% \begin{align} \nonumber
%     \bSigma_2  \!= \!\!\sum_{t=1}^T \!\bP_{t-1}^s \!+\! \bpsi_{t-1}^s {\bpsi^s}^\top_{t-1} \,,
%     \hspace{0.9ex}
%     \bSigma_3  \!= \!\!\sum_{t=1}^T \!\by_t {\bpsi_{t}^s}^\top,
%     \hspace{0.9ex}
%     \bSigma_5  \!=  \!\!\sum_{t=1}^T \!\by_t\by_t^\top,
% \end{align}
\begin{align}
\begin{split}
    \bSigma_1  = \sum_{t=1}^T \bP_{t}^s + \bpsi_{t}^s {\bpsi_{t}^s}^\top,
    \hspace{4.25ex}
    \bSigma_4  =  \sum_{t=1}^T \bP_{t}^s {\bG^s}^\top_{t-1} + \bpsi_{t}^s {\bpsi^s}^\top_{t-1}
\end{split}
\nonumber\\\nonumber
\begin{split}
    \bSigma_2  \!= \!\!\sum_{t=1}^T \!\bP_{t-1}^s \!+\! \bpsi_{t-1}^s {\bpsi^s}^\top_{t-1} \,,
    \hspace{0.9ex}
    \bSigma_3  \!= \!\!\sum_{t=1}^T \!\by_t {\bpsi_{t}^s}^\top,
    \hspace{0.9ex}
    \bSigma_5  \!=  \!\!\sum_{t=1}^T \!\by_t\by_t^\top,
\end{split}
\end{align}

Under the assumption that $\bR=\sigma_r^2\bI_{NL}$, optimizing $\mathcal{Q}(\btheta,\btheta^{(t)})$ with respect to $\bP_{0|0}$, $\bQ$, $\bR$ and $\bpsi_{0|0}$ is relatively straightforward and can be done as~\cite{sarkka2013bayesian}
\begin{align}
    %\bP_{0|0}^* & = \bP_0^s + (\bpsi_0^s-\bpsi_{0|0}(\btheta))(\bpsi_0^s-\bpsi_{0|0}(\btheta))^\top \label{eq:P00*}
    & \bP_{0|0}^* = \bP_0^s + (\bpsi_0^s-\bpsi_{0|0})(\bpsi_0^s-\bpsi_{0|0})^\top \label{eq:P00*}
    \\
    & \bQ^* = \bSigma_1 - \bSigma_4 - \bSigma_4^\top + \bSigma_2
    \\
    % & \sigma_r^* = (LN)^{-1}\tr\big\{\bSigma_5 - 2\bH(\!\bA) \bSigma_3^\top \!+\! \bH(\!\bA) \bSigma_1 \bH(\!\bA)^\top\big\}
    & \sigma_r^* = \tr\big\{\bSigma_5 - 2\bH(\!\bA) \bSigma_3^\top \!+\! \bH(\!\bA) \bSigma_1 \bH(\!\bA)^\top\big\}\big/(LN)
    \\
    & \bpsi_{0|0}^* = \bpsi_0^s. \label{eq:psi00*}
\end{align}

The optimization w.r.t. $\bH(\!\bA)$ is, however, more complex due to the structure of this matrix. Since $\bR=\sigma_r^2\bI$ for some $\sigma_r>0$, the problem can be stated as
% \begin{align} \label{eq:abundances_cost_function}
%     \min_{\bA} \,\,\, \tr\!\Big\{\bH(\bA) \bSigma_1 \bH(\bA)^\top - \bSigma_3 \bH(\bA)^\top - \bH(\bA)\bSigma_3^\top \Big\}
% \end{align}
\begin{align} \label{eq:abundances_cost_function}
    \widehat{\!\bA} = \mathop{\arg\min}_{\bA} \,\, \tr\!\Big\{\bH(\bA) \bSigma_1 \bH(\bA)^\top - 2\,\bSigma_3 \bH(\bA)^\top \Big\}
\end{align}
% Note that, since $\bH(\bA)$ is a linear transformation of $\bA$, the cost function in~\eqref{eq:abundances_cost_function} is quadratic and convex in $\bA$, and can be solved efficiently. The first term of~\eqref{eq:abundances_cost_function} can be written as:
%
In order to solve~\eqref{eq:abundances_cost_function} efficiently, we rewrite its terms in the following to explore the structure of $\bH(\bA)$. For the first term:
\begin{align}%\small
    \tr\{\bH(\bA) \bSigma_1 \bH(\bA)^\top\} 
    % {}={} & \tr\{\bH(\bA)^\top\bH(\bA) \bSigma_1 \} \nonumber
    % \\ 
    {}={} & \tr\{(\bA\otimes\bI_L)(\bA^\top\otimes\bI_L)\widetilde{\bSigma}_1\}
    \nonumber\\
    {}={} & \tr\{(\bA\bA^\top\otimes\bI_L)\widetilde{\bSigma}_1\}
\end{align}
% \vspace{-0.3cm}
\noindent where $\widetilde{\bSigma}_1=\diag(\bm_0)\bSigma_1\diag(\bm_0)$. To explore the properties of the Kronecker product and simplify the solution to this problem, we assume that $\widetilde{\bSigma}_1$ can be decomposed as~\cite{van1993approximationKronecker,batselier2017constructiveKroneckerTensors} 
$
\widetilde{\bSigma}_1 = \sum_{k=1}^{K_1} \bC_k \otimes \bD_k, \quad \bC_k \in\amsmathbb{R}^{P\times P} , \quad \bD_k \in\amsmathbb{R}^{L\times L},
$
% \begin{align} \label{eq:kronecker_dec1a}
%     \widetilde{\bSigma}_1 = \sum_{k=1}^{K_1} \bC_k \otimes \bD_k,
%     \quad \bC_k \in\amsmathbb{R}^{P\times P} , \quad \bD_k \in\amsmathbb{R}^{L\times L}
% \end{align}
and using the properties of the Kronecker product, we have
\begin{align}  \label{eq:kronecker_dec1b}
    \tr\{(\bA\bA^\top\otimes\bI_L)\widetilde{\bSigma}_1\} 
    % = & \sum_{k=1}^{K_1} \tr\{(\bA\bA^\top\otimes\bI_L)(\bC_k \otimes \bD_k)\}
    % \\
    % = & \sum_{k=1}^{K_1} \tr\{(\bA\bA^\top\bC_k) \otimes \bD_k \}
    % \\
    = & \sum\nolimits_{k=1}^{K_1} \tr\{(\bA\bA^\top\bC_k)\} \tr\{\bD_k\} \,.
\end{align}

Similarly, for the second term, we have 
$
\tr\{\bH(\bA)\bSigma_3^\top\} =  \tr\{(\bA^\top\otimes\bI_L)\widetilde{\bSigma}_3^{\top}\}
$
% \begin{align}
%     \tr\{\bH(\bA)\bSigma_3^\top\} = & \tr\{(\bA^\top\otimes\bI_L)\widetilde{\bSigma}_3^{\top}\}
% \end{align}
where $\widetilde{\bSigma}_3=\bSigma_3 \diag(\bm_0)$. By decomposing $\widetilde{\bSigma}_3$ as in~\cite{van1993approximationKronecker,batselier2017constructiveKroneckerTensors} leads to 
$
\widetilde{\bSigma}_3 = \sum_{k=1}^{K_2} \widetilde{\bC}_k \otimes \widetilde{\bD}_k,
    \quad \widetilde{\bC}_k \in\amsmathbb{R}^{N\times P} , \quad \widetilde{\bD}_k \in\amsmathbb{R}^{L\times L},
$
% \begin{align} \label{eq:kronecker_dec2a}
%     \widetilde{\bSigma}_3 = \sum_{k=1}^{K_2} \widetilde{\bC}_k \otimes \widetilde{\bD}_k,
%     \quad \widetilde{\bC}_k \in\amsmathbb{R}^{N\times P} , \quad \widetilde{\bD}_k \in\amsmathbb{R}^{L\times L}
% \end{align}
and using the properties of the Kronecker product, we have
\begin{align} \label{eq:kronecker_dec2b}
    \tr\{(\bA^\top\otimes\bI_L)\widetilde{\bSigma}_3\} 
    % = & \sum_{k=1}^{K_2} \tr\{(\bA^\top\otimes\bI_L)(\widetilde{\bC}_k^\top \otimes \widetilde{\bD}_k^\top)\}
    % \\
    % = & \sum_{k=1}^{K_2} \tr\{(\bA^\top\widetilde{\bC}_k^\top) \otimes \widetilde{\bD}_k^\top\}
    % \\
    = & \sum\nolimits_{k=1}^{K_2} \tr\{(\bA^\top\widetilde{\bC}_k^\top)\} \tr\{\widetilde{\bD}_k^\top\} \,.
\end{align}
By substituting \eqref{eq:kronecker_dec1b} and \eqref{eq:kronecker_dec2b} in~\eqref{eq:abundances_cost_function}, taking the derivative of the cost function with respect to $\bA$ and setting it equal to zero, we obtain the following solution for $\bA$:
% Using these results, the derivative of the cost function with respect to $\bA$ can be obtained as:
% \begin{align}
% \cblue{J_{\bA}} = \sum_{k=1}^{K_1} \tr\{\bD_k\} (\bC_k\bA + \bC_k^\top\bA) 
%     - 2 \sum_{k=1}^{K_2} \tr\{\widetilde{\bD}_k^\top\} \widetilde{\bC}_k^\top
%     % \nabla_{\bA}Cost = \sum_{k=1}^{K_1} \tr\{\bD_k\} (\bC_k\bA + \bC_k^\top\bA) 
%     % - 2 \sum_{k=1}^{K_2} \tr\{\widetilde{\bD}_k^\top\} \widetilde{\bC}_k^\top
% \end{align}
% which leads to the following solution for $\bA$:
% \begin{align}\label{eq:sol_A}
%     {\bA}^* = \bigg[\sum\nolimits_{k=1}^{K_1} \tr\{\bD_k\} (\bC_k + \bC_k^\top) \bigg]^{-1} \bigg[2\sum\nolimits_{k=1}^{K_2} \tr\{\widetilde{\bD}_k^\top\} \widetilde{\bC}_k^\top \bigg].
% \end{align}
\begin{align}\label{eq:sol_A}
    \!\!\! \widehat{\!\bA} \!= \!\Big[\!\sum\nolimits_{k=1}^{K_1} \!\!\tr\{\bD_k\} (\bC_k \!+ \bC_k^\top) \Big]^{-1} \Big[2\!\sum\nolimits_{k=1}^{K_2} \!\!\tr\{\widetilde{\bD}_k^\top\} \widetilde{\bC}_k^\top \Big].
\end{align}

Although the approach presented in sections~\ref{sec:kalmanFilt}--~\ref{sec:expectM} provides an estimate~$\,\widehat{\!\bA}$ of the average abundances, the temporal abundance variations $\Delta\bA_t$ can make~$\,\widehat{\!\bA}$ an inaccurate approximation of $\bA_t$ for some image sequences (e.g. when sudden changes are present). To mitigate this issue, we compensate the abundance variations $\Delta\bA_t$ by solving using a fully constrained least squares (FCLS) problem:
\begin{align} \label{eq:final_reg_fcls_A}
\begin{split}
    & \mathop{\min}_{\bA_{t}} \,\,
    \|\bY_{\!t} - (\bM_0\odot\widehat{\bPsi}_t)\bA_{t}\|_F^2 + \lambda\|\bA_{t}-\,\widehat{\!\bA}\|_F^2 
    \\[-0.2cm]
    & \text{ s.t. } \,\,  \bA_{t} \geq0, \, \cb{1}^\top\bA_{t} = \cb{1}^\top,
\end{split}
\end{align}
for $t=1,\ldots,T$, where $\widehat{\bPsi}_t$ is the matrix-ordered version of the estimated states $\bpsi_t^s$ and $\lambda\geq0$ is a regularization parameter. 
% This gives the final abundance estimate for each time instant.
The proposed methodology is summarized in Algorithm~\ref{alg:algorithm}.

% \todo[inline]{It is not clear how you combine $\widehat{\!\bA}$ obtained from \eqref{eq:sol_A} with the solution of the FCLS problem to obtain the final solution $\bA_t^*$. From the text and from the algorithm it looks like you simply do not use the solution from \eqref{eq:sol_A} outside the for loop.  }

% \todo[inline]{In the algorithm, it is not clear the sequence of calculations.  The way it is written it looks like the initializations are determined after using the to solve (9)-(11).  Should the steps in the algorithm be indexed in time? Please clarify?}

% The proposed methodology is summarized in Algorithm~\ref{alg:algorithm}, namely, \emph{Dynamical Spectral Unmixing with Kalman Smoother} (DySUKS) \cblue{[we need a better name]}.  

% \vspace{-0.3cm}
\begin{algorithm} [bth!]
\footnotesize
% \scriptsize
\SetKwInOut{Input}{Input}
\SetKwInOut{Output}{Output}
\caption{\mbox{Kalman filter and Smoother for MTSU}\label{alg:algorithm}}
\Input{$\big\{\by_t\big\}_{t=1}^T$, $\bA^{(0)}$, $\bpsi^{(0)}$, $\bM_0$. $\bpsi_{0|0}^{(0)}$, $\bQ^{(0)}$, $\sigma_r^{(0)}$, $\bP_{0|0}^{(0)}$, $\lambda$}
% \Output{$\bA_t^*$, $\bM_t^*$.}

\For{$i=1,\ldots,K_{\max}$}{
    Estimate $\bpsi_t$ using~\eqref{eq:kalman_filt_pred}--\eqref{kalman_filt_update} for $t=1,\ldots,T$\;
    Estimate $\bpsi^{s}_t$ and $\bP_t^s$ using~\eqref{eq:kalman_smoother} for $t=T,\ldots,1$\;
    % Estimate the average abundances $\bA^{(i)}$ using~\eqref{eq:sol_A}\;
    % Estimate $\bP^*_{0|0},\, \bQ^*, \sigma_r^*$ and $\bpsi^*_{0|0}$ using~\eqref{eq:P00*}--\eqref{eq:psi00*}\;
    Estimate $\bP_{0|0}^{(i)},\, \bQ^{(i)}, \sigma_r^{(i)}$, $\bpsi_{0|0}^{(i)}$, $\bA^{(i)}$ using~\eqref{eq:P00*}--\eqref{eq:psi00*},~\eqref{eq:sol_A}\; %\eqref{eq:abundances_cost_function}\;
}%\EndFor
Estimate the temporal abundance variations according to~\eqref{eq:final_reg_fcls_A}\;
\KwRet $\bA_t^*=\,\widehat{\!\bA}_t$,~ $\bM_t^*=\bM_0 \odot \vect^{-1}(\bpsi^s_t)$, for $t=1,\ldots,T$
% $\{\bpsi^*_t\}_{t=1}^{T}=\{\bpsi^s_t\}_{t=1}^{T}$ \;
\end{algorithm}

\section{Experimental Results} \label{sec:Simulations}

In this section we evaluate the performance of the proposed method by comparing it with the fully constrained least squares (FCLS), and with the Online Unmixing (OU) strategy proposed in~\cite{thouvenin2016online}. \cred{To illustrate how the use of temporal information improves SU, we have also applied the GLMM algorithm~\cite{imbiriba2018GLMM} (which considers SV only within a single HI) to process each HI independently.}
In all experiments, the reference EME matrix $\bM_0$ was extracted from the observed HI at $t=1$ using the VCA algorithm~\cite{Nascimento2005}, and the abundances were initialized with the corresponding FCLS result. The other parameters were initialized as $\bpsi_{0|0}=\cb{1}$, $\bQ=0.1\bI$, $\sigma_r=0.01$, $\bP_{0|0}=\bI$ and $\lambda=10^{-8}$, and five EM iterations were considered.
The parameters of the OU algorithm were searched in the ranges detailed in the original publication~\cite{thouvenin2016online}.

The performance of the methods is evaluated using the average Normalized Root Mean Squared Error (NRMSE) between the estimated abundances ($\text{NRMSE}_{\bA}$), endmembers ($\text{NRMSE}_{\bM}$) and between the reconstructed HIs. NRMSE is defined as
$
\text{NRMSE}_{\bX} = \frac{1}{T} \sum_{t=1}^T \sqrt{\|\bX_{\!t} - \bX_{\!t}^*\|^2_F \,\,/\,\, \|\bX_{\!t}\|^2_F}
$,
% \begin{equation}\footnotesize
% \text{NRMSE}_{\bX} = \frac{1}{T} \sum_{t=1}^T \sqrt{\|\bX_{\!t} - \bX_{\!t}^*\|^2_F \,\,\Big/\,\, \|\bX_{\!t}\|^2_F}
% \end{equation}
where $\bX_{\!t}$ and $\bX_{\!t}^*$ are a true and an estimated variable, respectively, at time instant $t$. We also consider the average Spectral Angle Mapper (SAM) between the estimated endmembers, defined as
$
\text{SAM}_{\bM} = \frac{1}{T}\sum_{t=1}^{T}\sum_{k=1}^{P}\arccos (\bm_{k,t}^\top\bm_{k,t}^*/\|\bm_{k,t}\|\|\bm_{k,t}^*\|).
$
\begin{table}[htb!]
\footnotesize
% \scriptsize
% \small
\caption{Quantitative simulation results (values $\times100$).}
\vspace{-0.3cm}
\centering
\renewcommand{\arraystretch}{1}
\setlength{\tabcolsep}{3.3pt}
% \vspace{-0.5cm}
\begin{tabular}{l|cccc|c}
% \toprule
% \bottomrule
\midrule
& \multicolumn{4}{|c|}{\cred{Synthetic Data (average results)}} & \cred{Real Data} \\
% \toprule%\bottomrule
\midrule
& $\text{NRMSE}_{\bA}$ & $\text{NRMSE}_{\bM}$ &$\text{SAM}_{\bM}$ & $\text{NRMSE}_{\bY}$ &  \cred{$\text{NRMSE}_{\bY}$} \\ \midrule
FCLS     	&	4.60	&	3.30	&	2.47	&	4.03	&   8.73\\
\cred{GLMM} &   4.87    &   4.25    &   3.69 &   \textbf{2.93}    &   \textbf{0.15}\\
OU       	&	3.40	&	2.82	&	2.05	&	3.09	&   3.48\\
Proposed 	&	\textbf{2.65}	&	\textbf{2.06}	&	\textbf{1.83}	&	3.12	&   9.02\\
\bottomrule%\toprule
\end{tabular}
\label{tab:quantitativeResults}
\end{table}

% \vspace{-0.5cm}
% \subsection{Synthetic data}
% \subsection{Experiments}

A synthetic dataset with $L=173$ bands, $N=50$ pixels and $T=10$ frames was created by generating abundance values sampled from a Dirichlet distribution. HIs containing three endmembers (vegetation, water and soil) were generated following the model~\eqref{eq:model_glmm} to generate one different EME matrix for each time instant.
% \todo[inline]{Do the endmembers in the GLMM model include spatial variability? If the example includes both spatial and time variability this should be emphasized.}
%
Temporal spectral variability was introduced by performing a random walk according to the model $\bpsi_t=\bF\bpsi_{t-1}+\cb{q}_t$ with $\bF=0.9\bI$, with $\bpsi_0=\cb{1}_{LP}$ and covariance $0.01\bI$. 
Finally, white Gaussian noise was added to the images, resulting in an SNR of 30dB. 
The $\text{SAM}_{\bM}$ metric for the GLMM method was computed by considering the average endmember matrix (across all pixels) for each time instant.
%
% In order to evaluate the performance of the algorithms, we performed 900 Monte Carlo runs. Additionally, we also simulated different amounts of temporal abundance variability, with an average (per pixel) temporal standard deviation of approximately~$3\times10^{-3}$. \cblue{[Jose: What is the meaning of ``per pixel average temporal standard deviation"? Is the average evaluated in time? Is the standard deviation different for each time instant? Is it different for different pixels at the same time instant?]}
In order to evaluate the performance of the algorithms, we performed 900 Monte Carlo runs. Additionally, we also simulated different amounts of temporal abundance variability, with the temporal standard deviation of each pixel being, on average, approximately~$3\times10^{-3}$.
%*100=  0.0045    0.0100    0.0142    0.0316    0.0447    0.0987    0.1407    0.3160    0.4485

The average metrics (across all MC runs and variance values) are shown in Table~\ref{tab:quantitativeResults}. It can be seen that the proposed method performs significantly better than the other algorithms.
Improvements over OU can be found in $\text{NRMSE}_{\bA}$ ($22\%$), $\text{NRMSE}_{\bM}$ ($27\%$) and $\text{SAM}_{\bM}$ ($11\%$). The GLMM provided the smallest reconstruction error $\text{NRMSE}_{\bY}$ since it has the largest amount of degrees of freedom. However, this did not translate into better abundance or EME estimates, since spatial SV was not present in the data and time information was not taken into account.

\begin{figure}[!t]
    \centering
    \includegraphics[width=0.8\linewidth]{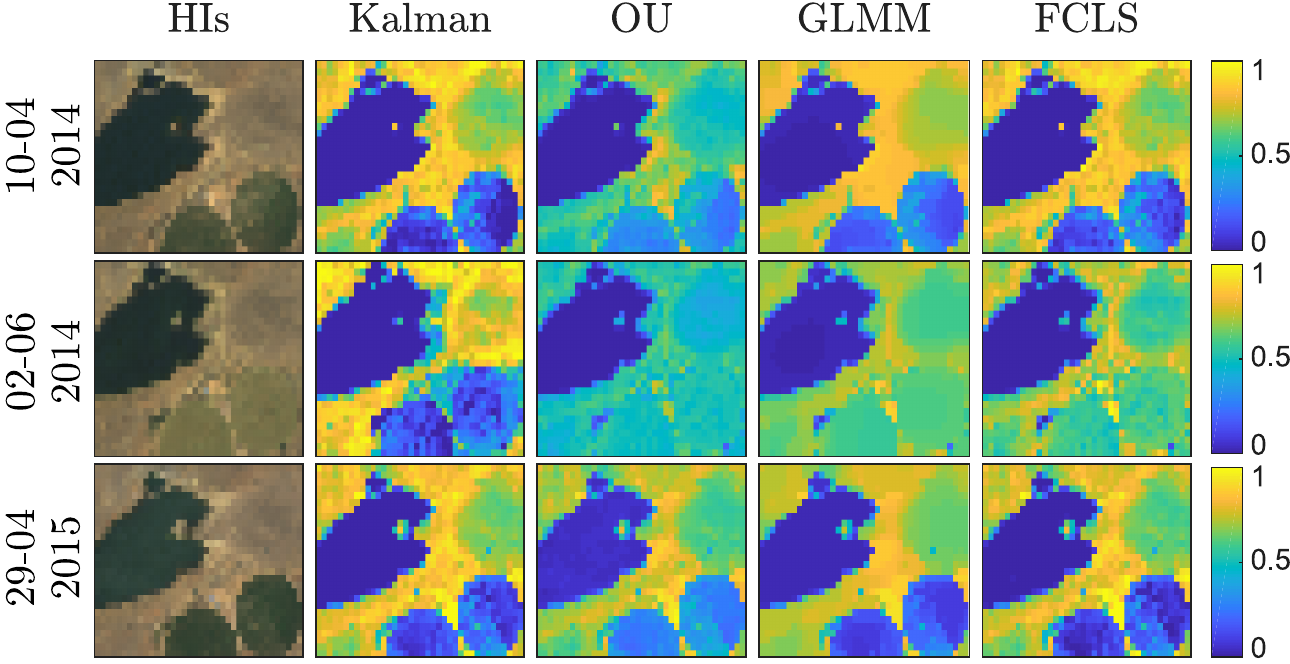}
    \vspace{-0.3cm}
    \caption{Abundance maps for the Soil endmember (right) and visual representation (left) of the Lake Tahoe HI sequence.} %\cblue{[Jose: Shouldn't we add a color scale?]}}
    \label{fig:realIm}
    % \vspace{-0.5cm}
\end{figure}

%
%
% \subsection{Real data}
%
%

% https://pthouvenin.github.io/robust-unmixing-plmm/
% https://pthouvenin.github.io/papers/Thouvenin_EUSIPCO_2017.pdf
% https://pthouvenin.github.io/papers/Thouvenin_IEEE_ICASSP_2016.pdf
% https://github.com/pthouvenin/robust-unmixing-plmm
% https://hal.archives-ouvertes.fr/hal-01363316/file/halimi_15346.pdf

For the simulations with real data, we consider three images from the Lake Tahoe sequence, originally studied in~\cite{thouvenin2016online}\cred{, acquired at 10-04-2014, 02-06-2014 and at 29-04-2015}. \cred{These images can be seen in Fig.~\ref{fig:realIm}, and are composed by three EMEs, water (a lake), soil and vegetation (two crop circles, whose aspect varies considerably between the three HIs).} The images were first downsized to $28\times 38$ pixels for faster processing, and contained $L=173$ bands. $\bM_0$ and the OU EME initialization were constructed using the same signatures as in~\cite{thouvenin2016online}. The OU parameters were the same as those used in~\cite{thouvenin2016online}. The results can be seen in Fig.~\ref{fig:realIm} \cred{and Table~\ref{tab:quantitativeResults}}. Due to space limitations, only the soil endmember is shown \cred{(more results can be found in the supplementary document, also available in~\cite{borsoi2020kalmanEM_arxiv})}. It can be seen that the proposed algorithm provides higher abundance values in the regions corresponding to soil in the HI, with significantly less confusion in the vegetation endmember when compared to the other methods, especially for $t=2$. Similar improvements could be noticed for the other EMEs as well.
\cred{The reconstruction error $\text{NRMSE}_{\bY}$ of the algorithms is generally related to their overall amount of degrees of freedom, thus being much larger for both the proposed method and for FCLS than for GLMM.}

%The reconstruction error $\text{NRMSE}_{\bY}$, shown in Table~\ref{tab:quantitativeResults}, shows that the proposed method was not able to represent the HIs as well as the other methods, which is likely due to the higher temporal variation of the abundances in this image.

% \vspace{-0.3cm}
\section{Conclusions}\label{sec:conclusions}

We proposed a new multitemporal spectral unmixing algorithm accounting for spectral variability. A state-space dynamical model was proposed for the time evolution of the coefficients encoding the spectral variability of the endmembers. Bayesian filtering was used to estimate the state variables. Assuming small abundance variations in short time intervals, expectation maximization employed to efficiently estimate the remaining parameters, including the fractional abundances. Simulation results indicate that the proposed method can outperform state-of-the-art multitemporal spectral unmixing algorithms. A future perspective is the extension of the method to properly handle abrupt abundance changes.

% \clearpage 

% \clearpage
\bibliographystyle{IEEEbib}
% \bibliography{../bibliography,references,bibliography}
\bibliography{bibliography}

\onecolumn

\centerline{{\huge Supplemental Material:}}
\medskip
\centerline{{\huge Kalman Filtering and Expectation Maximization}}
\medskip
\centerline{{\huge for Multitemporal Spectral Unmixing}}
% \bigskip
% The following material supplements the paper 

% \bigskip
% \noindent [A] R. A. Borsoi, T. Imbiriba, J. C. M. Bermudez and C. Richard, ``A Fast Multiscale Spatial Regularization for Sparse Hyperspectral Unmixing'', \cred{[complete reference after publication]}.

% \setcounter{section}{0}
% \setcounter{table}{0}
% \setcounter{figure}{0}

\section{Supplemental Material: Further discussion about the observation noise}

Note that the observation noise $\br_{t}$ in~(7) is correlated with the state $\bpsi_{t}$, which means that traditional algorithms such as the Kalman filter are no longer statistically optimal. This problem is similar to the linearization error observed in the extended Kalman filter for nonlinear models. Although state-dependent noise has been considered in linear state estimation, a proper treatment involves $\mathcal{O}(N^4L^4)$ operations per time instant, which is computationally intractable~[S1].
Fortunately, previous work~[S2] indicates that even if signal-dependent noise is present in SU applications, the use of signal-independent noise approximations still provides competitive performance. Using this knowledge, in  this work we ignore this dependence and apply the Kalman filter and smoother to estimate $\{\bpsi_t\}$. This should work satisfactorily as long as the contribution of $\big(\bDelta\!\bA_t^\top\otimes\bI_L\big)\diag(\bm_{0}) \bpsi_{t}$ in~(7) is not too large.
%Fortunately, previous work indicate that even if signal-dependent noise is present in SU applications, the use of signal-independent noise approximations still provides competitive performance~\cite{ammanouil2014blind}.

% \bibliographystyle{IEEEtran}
% %\bibliography{references,references_nonlSppx,ourpapers}
% \bibliography{bibliography}

\section{Supplemental Material: Additional simulation results}

\begin{figure}[!h]
    \centering
    \includegraphics[width=0.7\linewidth]{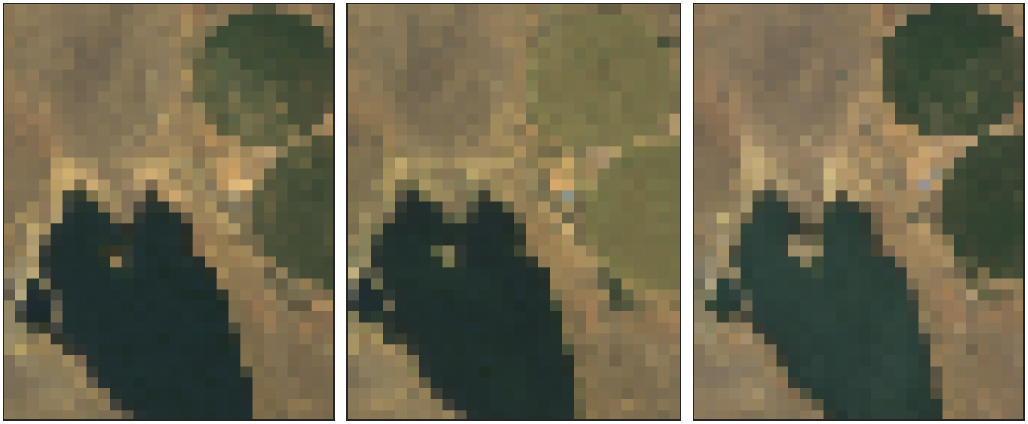}
    \caption{Visual representation of the Lake Tahoe dataset (acquisition dates are 10-04-2014, 02-06-2014 and 29-04-2015, for the left, center, and right images, respectively).}
    \label{fig:tahoe_HIs}
\end{figure}

\begin{figure}
    \centering
    \includegraphics[width=0.7\linewidth]{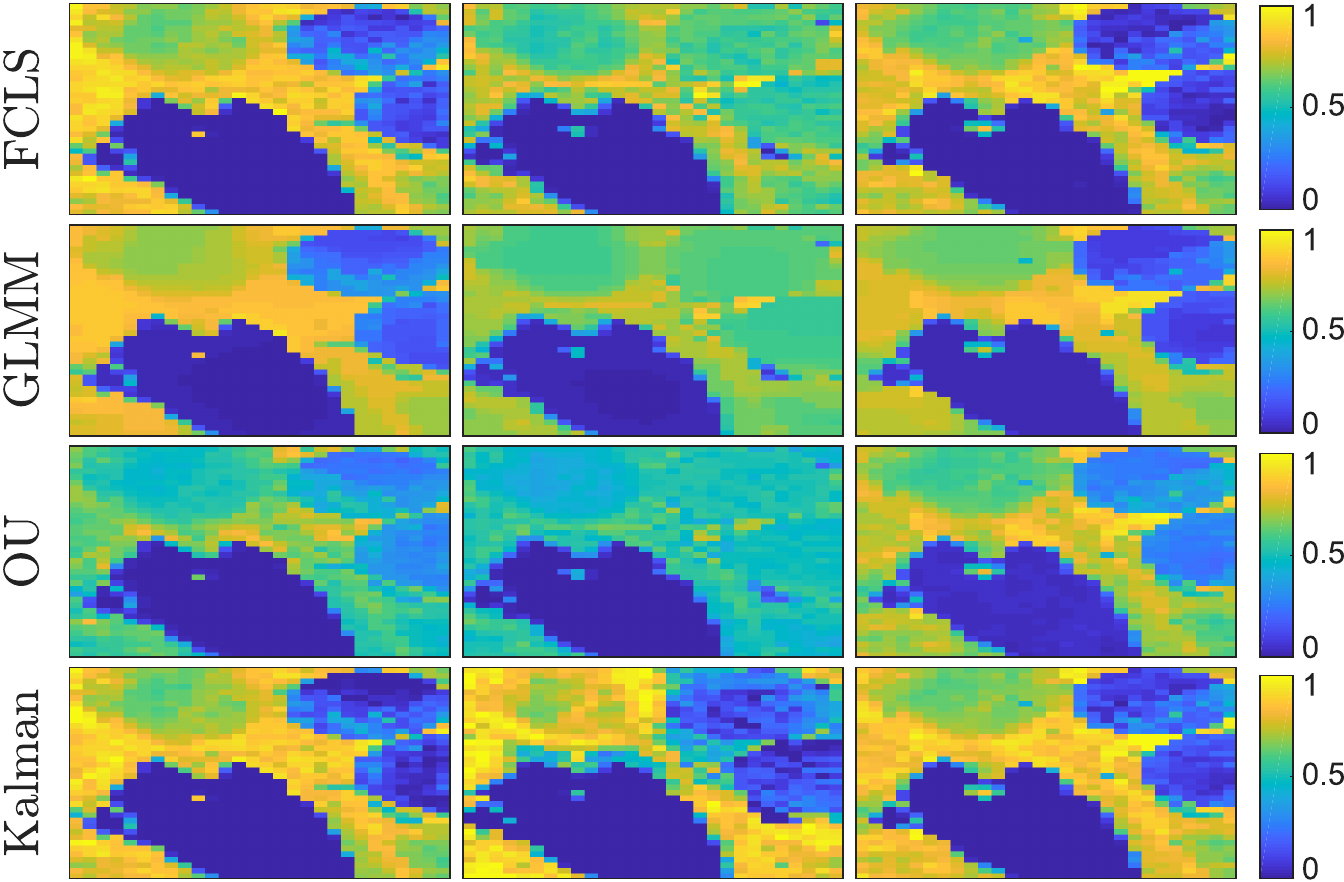}
    \caption{Abundance maps of the Soil endmember from the Lake Tahoe dataset.}
    \label{fig:real_soil}
\end{figure}

\begin{figure}
    \centering
    \includegraphics[width=0.7\linewidth]{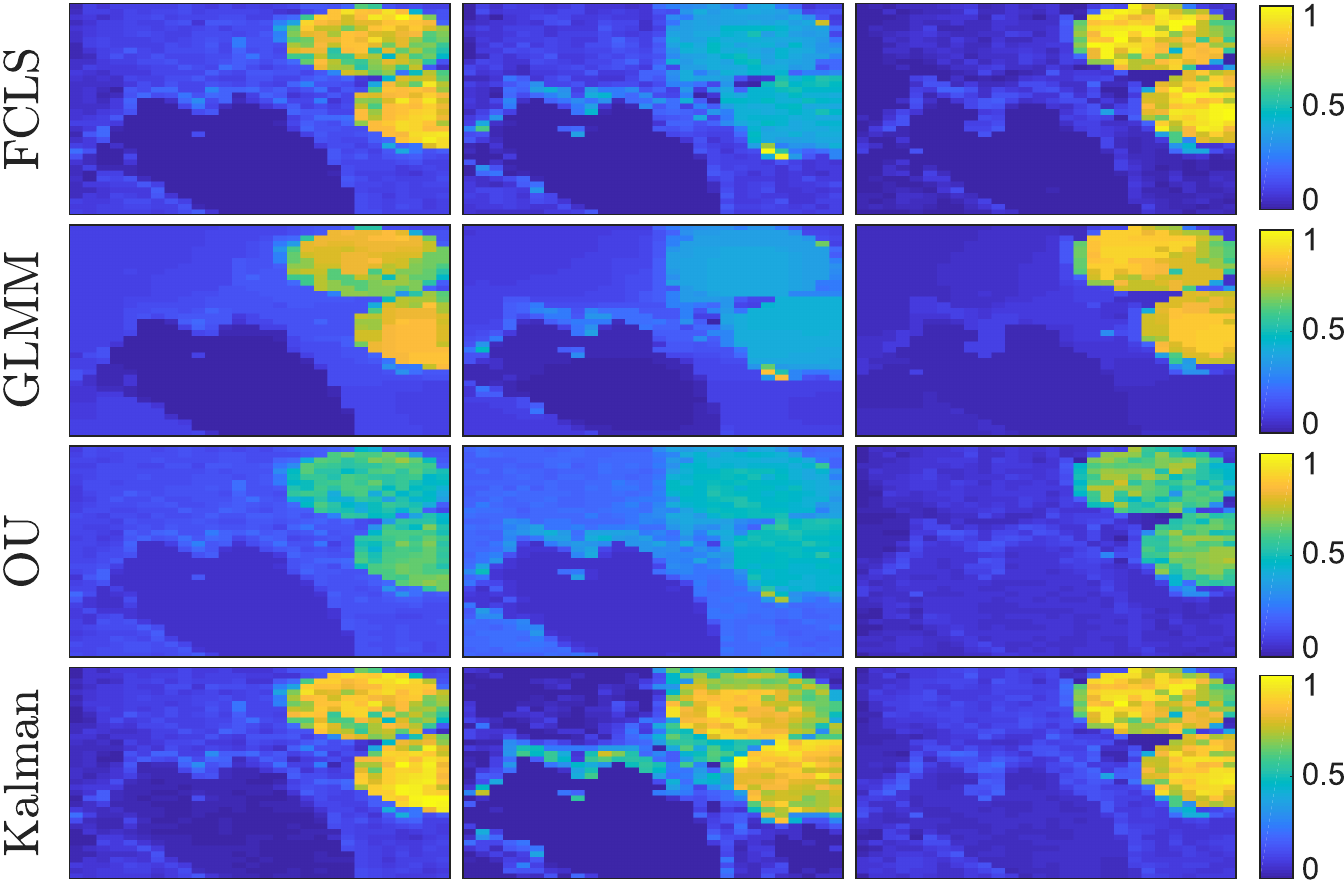}
    \caption{Abundance maps of the Vegetation endmember from the Lake Tahoe dataset.}
    \label{fig:real_veg}
\end{figure}

\begin{figure}
    \centering
    \includegraphics[width=0.7\linewidth]{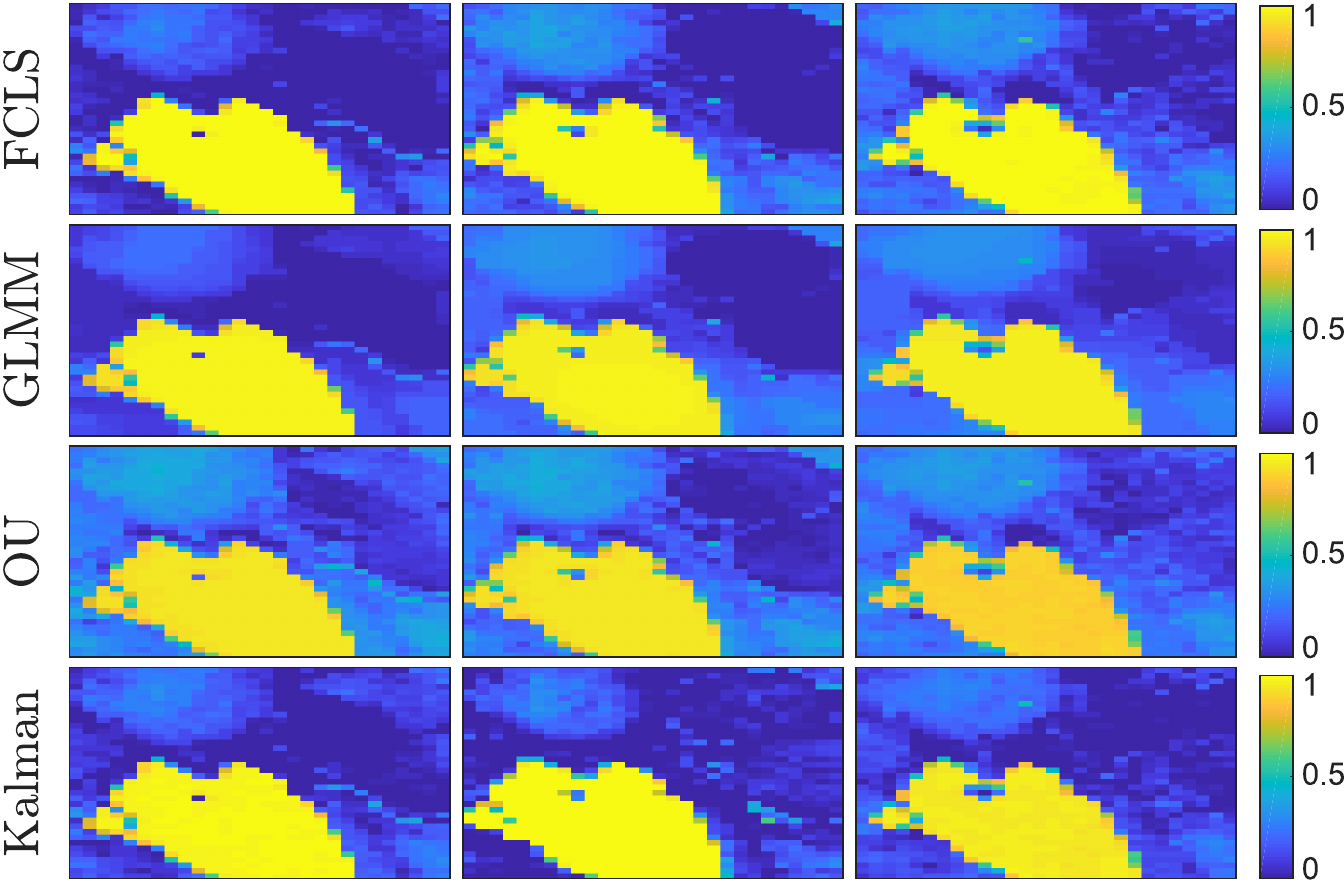}
    \caption{Abundance maps of the Water endmember from the Lake Tahoe dataset.}
    \label{fig:real_water}
\end{figure}

\begin{figure}
    \centering
    \includegraphics[width=0.9\linewidth]{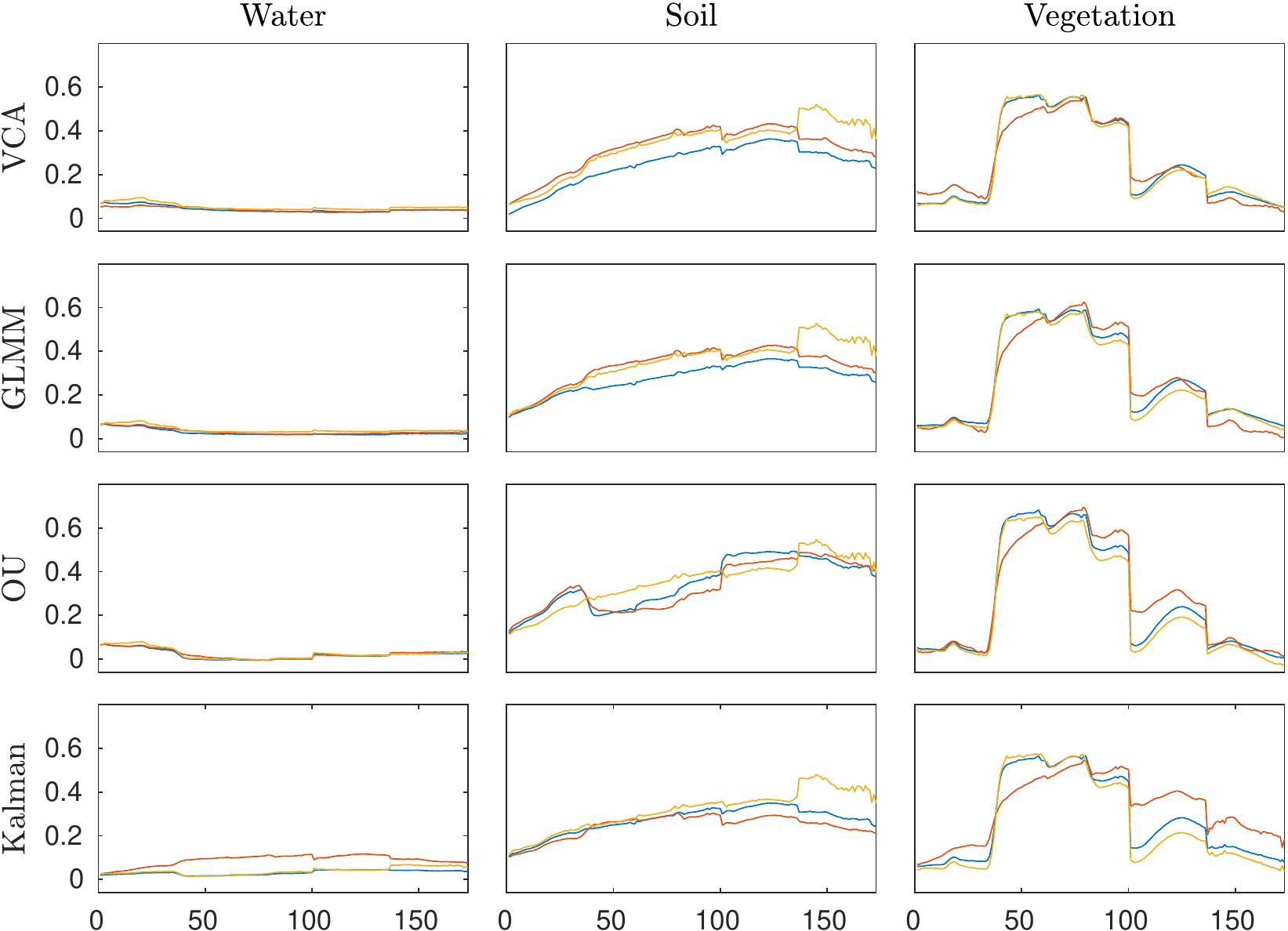}
    \caption{Estimated endmembers for the Lake Tahoe dataset (blue, red and yellow curves correspond to the left, center, and right images, respectively).}
    \label{fig:real_endmembers}
\end{figure}

Here we present additional visual results corresponding to the Lake Tahoe dataset. A true color representation of the observed HIs can be seen in Fig.~ \ref{fig:tahoe_HIs}. The estimated abundance maps of the Soil, Vegetation and Water endmembers are shown in Figs.~\ref{fig:real_soil},~\ref{fig:real_veg} and~\ref{fig:real_water}. Finally, the spectral signatures of the recovered endmembers are shown in Fig.~\ref{fig:real_endmembers}.

It can be seen that even though only three HIs were contained in the input image sequence, the Kalman-based solution was able to obtain good estimates for both the endmembers and abundance maps, with the estimated spatial distribution of the three materials in the scene agreeing well with a visual inspection of the HIs.

% \begin{thebibliography}{99}
%   \bibitem{Somers12} Somers, B., Zortea, M., Plaza, A., and Asner, G. P.. \textbf{Automated extraction of image-based endmember bundles for improved spectral unmixing}. IEEE Journal of Selected Topics in Applied Earth Observations and Remote Sensing, 5(2), 396-408, 2012.
  
% \end{thebibliography}

\section*{References}

[S1] D. Spinello and D. J. Stilwell, “Nonlinear  estimation  with state-dependent Gaussian observation noise,”IEEE Transactions on Automatic Control, vol. 55, no. 6, pp. 1358–1366, 2010.

[S2] R. Ammanouil, A. Ferrari, C. Richard, and D. Mary, “Blind and fully constrained unmixing of hyperspectral images,” IEEE Transactions on Image Processing, vol. 23, no. 12, pp. 5510–5518, 2014.

\end{document}